%% file: main.tex
\documentclass[sigconf]{acmart}

\usepackage{booktabs} % For formal tables
\usepackage{multirow}
\usepackage{subfigure}
\usepackage{verbatim}
\usepackage{CJKutf8}
\usepackage{flushend}

\usepackage{makecell}
\usepackage[utf8]{inputenc}
\setcopyright{rightsretained}
\acmDOI{10.475/123_4}

\acmISBN{123-4567-24-567/08/06}

\acmConference[SIGIR 2021]{The 44th International ACM SIGIR Conference
on Research and Development in Information Retrieval}{July 2021}{Online}
\acmYear{2021}
\copyrightyear{2021}
\acmArticle{4}
\acmPrice{15.00}

\editor{Jennifer B. Sartor}
\editor{Theo D'Hondt}
\editor{Wolfgang De Meuter}

\begin{document}

\title{Empowering News Recommendation with\\ Pre-trained Language Models}

\fancyhead{}

%\subtitle{FedCTR}
%\subtitlenote{The full version of the author's guide is available as
%  \texttt{acmart.pdf} document}

\author{Chuhan Wu$^1$, Fangzhao Wu$^2$, Tao Qi$^1$, Yongfeng Huang$^1$}

\affiliation{%
  \institution{$^1$Department of Electronic Engineering \& BNRist, Tsinghua University, Beijing 100084 \\ $^2$Microsoft Research Asia, Beijing 100080, China}
} 
\email{{wuchuhan15,wufangzhao,taoqi.qt}@gmail.com,yfhuang@tsinghua.edu.cn}

% \author{Chuhan Wu}
% \affiliation{%
%   \institution{Tsinghua University}
%     \city{Beijing}
%     \postcode{100084}
%   \state{China}
% }
% \email{wuchuhan15@gmail.com}

% \author{Fangzhao Wu}
% \affiliation{%
%   \institution{Microsoft Research Asia}
%   \city{Beijing}
%   \state{China}
%   \postcode{100080}
% }
% \email{wufangzhao@gmail.com}

% \author{Tao Qi}
% \affiliation{%
%   \institution{Tsinghua University}
%     \city{Beijing}
%     \postcode{100084}
%   \state{China}
% }
% \email{qit16@mails.tsinghua.edu.cn}
% \author{Heyuan Wang}
% \affiliation{%
%   \institution{Microsoft Research Asia}
%   \city{Beijing}
%   \state{China}
%   \postcode{100080}
% }
% \email{heyuanww@163.com}

% \author{Yongfeng Huang}
% \affiliation{%
%   \institution{Tsinghua University}
%     \city{Beijing}
%     \postcode{100084}
%   \state{China}
% }
% \email{yfhuang@tsinghua.edu.cn}

% \author{Xing Xie}
% \affiliation{%
%   \institution{Microsoft Research Asia}
%   \city{Beijing}
%   \state{China}
%   \postcode{100080}
% }
% \email{xing.xie@microsoft.com}

\begin{abstract}
Personalized news recommendation is an essential technique for online news services.
News articles usually contain rich textual content, and accurate news modeling is important for personalized news recommendation. 
Existing news recommendation methods mainly model news texts based on traditional text modeling methods, which is not optimal for mining the deep semantic information in news texts.
Pre-trained language models (PLMs) are powerful for natural language understanding, which has the potential for better news modeling.
However, there is no public report that show PLMs have been applied to news recommendation.
In this paper, we report our work on exploiting pre-trained language models to empower news recommendation. 
Offline experimental results on both monolingual and multilingual news recommendation datasets show that leveraging PLMs for news modeling can effectively improve the performance of news recommendation. 
Our PLM-empowered news recommendation models have been deployed to the Microsoft News platform, and achieved significant gains in terms of both click and pageview in both English-speaking and global markets.
\end{abstract}

%
% The code below should be generated by the tool at
% http://dl.acm.org/ccs.cfm
% Please copy and paste the code instead of the example below.
%




\maketitle

\input{data/introduction.tex}

\input{data/method.tex}
\input{data/experiment.tex}
\input{data/conclusion.tex}

\section*{Acknowledgments}
This work was supported by the National Natural Science Foundation of China under Grant numbers  U1936216 and U1936208.

\bibliographystyle{ACM-Reference-Format}
\bibliography{main}

\end{document}

%% file: data/introduction.tex
\section{Introduction}

News recommendation techniques have played critical roles in many online news platforms to alleviate the information overload of users~\cite{okura2017embedding}.
News modeling is an important step in news recommendation, because it is a core technique to understand the content of candidate news and a prerequisite for inferring user interests from clicked news.
Since news articles usually have rich textual information, news texts modeling is the key for understanding news content for news recommendation.
Existing news recommendation methods usually model news texts based on traditional NLP models~\cite{okura2017embedding,wang2018dkn,wu2019npa,wu2019nrms,wang2020fine,wu2020user,wu2020sentirec}.
For example,~\citet{wang2018dkn} proposed to use a knowledge-aware CNN network to learn news representations from embeddings of words and entities in news title.
~\citet{wu2019nrms} proposed to use multi-head self-attention network to learn news representations from news titles.
However, it is difficult for these shallow models to understand the deep semantic information in news texts~\cite{vaswani2017attention}.
In addition, their models are only learned from the supervisions in the news recommendation task, which may not be optimal for capturing the semantic information.

Pre-trained language models (PLMs) have achieved great success in NLP due to their strong ability in text modeling~\cite{devlin2019bert,liu2019roberta,dong2019unified,lan2019albert,yang2019xlnet,lample2019cross,bao2020unilmv2}.
Different from traditional models that are usually directly trained with labeled data in specific tasks, PLMs are usually first pre-trained on a large unlabeled corpus via self-supervision to encode universal text information~\cite{devlin2019bert}.
Thus, PLMs can usually provide better initial point for finetuning in downstream tasks~\cite{qiu2020pre}.
In addition, different from many traditional NLP methods with shallow models~\cite{kim2014convolutional,yang2016hierarchical,joulin2017bag,wu2018detecting}, PLMs are usually much deeper with a huge number of parameters.
For example, the BERT-Base model contains 12 Transformer layers and up to 109M parameters~\cite{devlin2019bert}.
Thus, PLMs may have greater ability in modeling the complicated contextual information in news text, which have the potentials to improve news text modeling for news recommendation.

In this paper, we present our work on empowering large-scale news recommendation with pre-trained language models.\footnote{Source codes will be available at https://github.com/wuch15/PLM4NewsRec.}
Different from existing news recommendation methods that use shallow NLP models for news modeling, we explore to model news with pre-trained language models and finetune them with the news recommendation task.
Offline experiments on real-world English and multilingual news recommendation datasets validate that incorporating PLMs into news modeling can consistently improve the news recommendation performance.
In addition, our PLM-empowered news recommendation models have been deployed to the Microsoft News platform.\footnote{https://microsoftnews.msn.com}
To our best knowledge, this is the first reported effort to empower large-scale news recommender systems with PLMs.
The online flight experiments show that our PLM-empowered news recommendation models achieved 8.53\% click and 2.63\% pageview gains in English-speaking markets, and 10.68\% click and 6.04\% pageview gains in other 43 global markets.

%% file: data/method.tex
\section{Methodology}\label{sec:Model}

In this section, we introduce the details of PLM-empowered news recommendation. 
We first introduce the general news recommendation model framework, and then introduce how to incorporate PLMs into this framework to empower news modeling.

\begin{figure}[!t]
  \centering
    \includegraphics[width=0.9\linewidth]{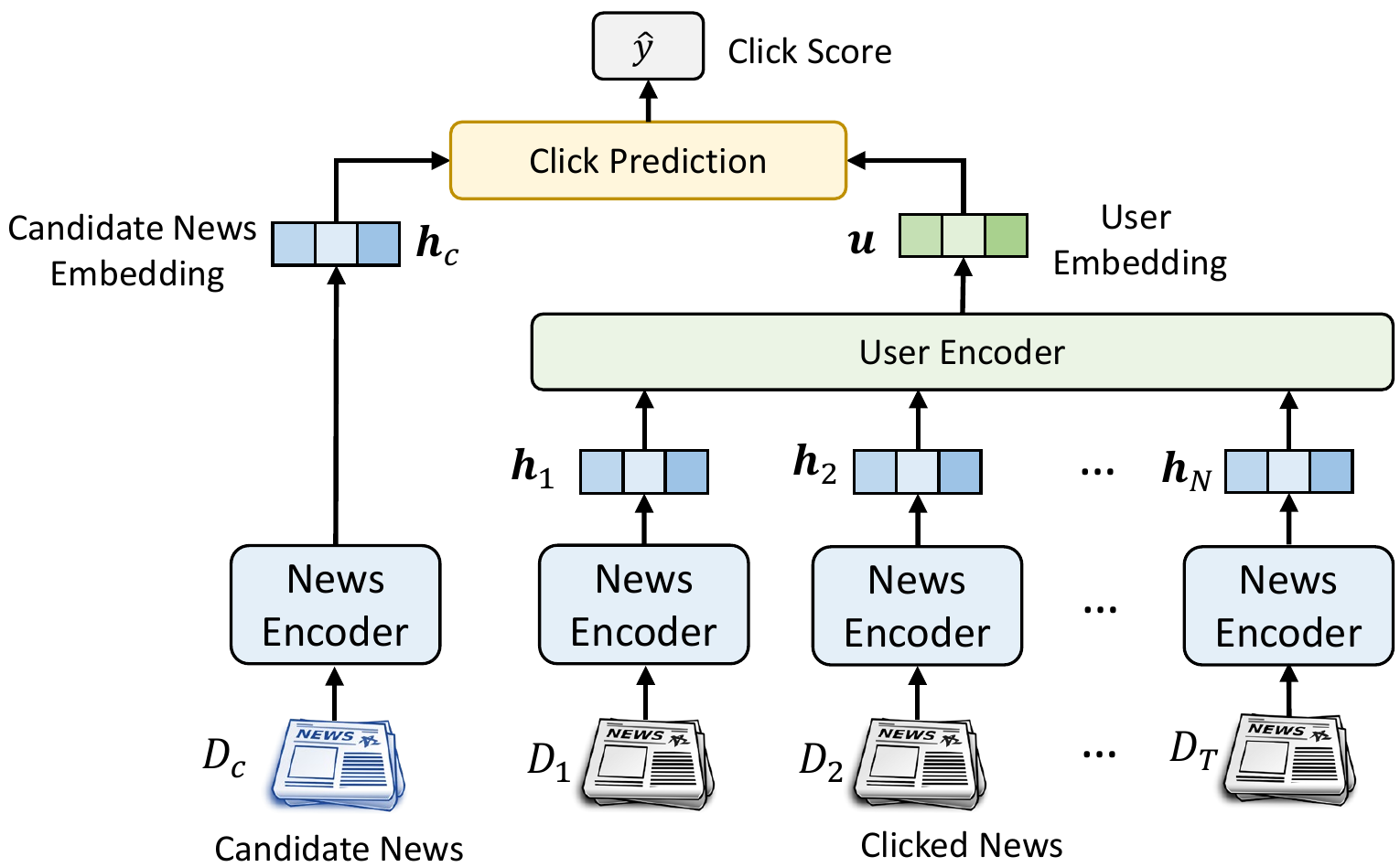}
  \caption{A common framework of news recommendation.}
  \label{fig.model0}
\end{figure}

\subsection{General News Recommendation Framework}

The general framework of news recommendation used in many existing methods~\cite{okura2017embedding,wu2019,wu2019nrms,an2019neural} is shown in Fig.~\ref{fig.model0}.
The core components in this framework include a news encoder that aims to learn the embeddings of news from their texts, a user encoder that learns user embedding from the embeddings of clicked news, and a click prediction module that computes personalized click score for news ranking based on the relevance between user embedding and candidate news embedding.
We assume a user has $T$ historical clicked news, which are denoted as $[D_1, D_2, ..., D_T]$.
The news encoder processes these clicked news of a user and each candidate news $D_c$ to obtain their embeddings, which are denoted as $[\mathbf{h}_1, \mathbf{h}_2, ..., \mathbf{h}_T]$ and $\mathbf{h}_c$, respectively.
It can be implemented by various NLP models, such as CNN~\cite{kim2014convolutional} and self-attention~\cite{vaswani2017attention}.
The user encoder receives the sequence of clicked news embeddings as input, and outputs a user embedding $\mathbf{u}$ that summarizes user interest information.
It can also be implemented by various models, such as the GRU network used in~\cite{okura2017embedding}, the attention network used in~\cite{wu2019} and the combination of multi-head self-attention and additive attention networks used in~\cite{wu2019nrms}.
The click prediction module takes the user embedding $\mathbf{u}$ and $\mathbf{h}_c$ as inputs, and compute the click score $\hat{y}$ by evaluating their relevance.
It can also be implemented by various methods such as inner product~\cite{okura2017embedding}, neural network~\cite{wang2018dkn} and factorization machine~\cite{guo2017deepfm}.
%Following prior works~\cite{okura2017embedding,wu2019npa,wu2019nrms}, we use inner product to implement this module, and the click score is computed by $\hat{y}=\mathbf{h}_c \cdot \mathbf{u}$.

\subsection{PLM Empowered News Recommendation}

\begin{figure}[!t]
  \centering
    \includegraphics[width=0.9\linewidth]{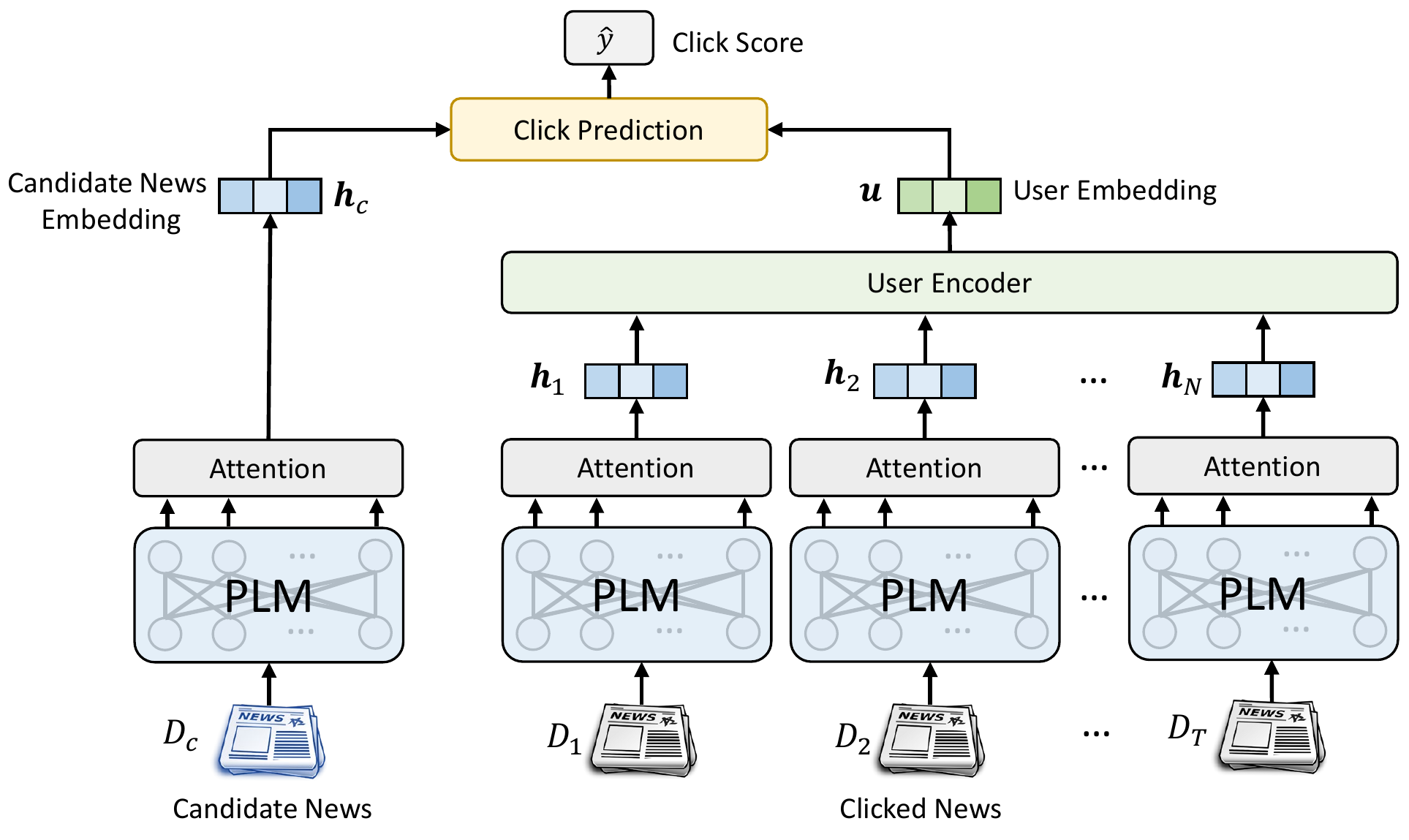}
  \caption{The framework of PLM empowered news recommendation.}
  \label{fig.model}
\end{figure}

Next, we introduce the framework of PLM empowered news recommendation, as shown in Fig.~\ref{fig.model}.
We instantiate the news encoder with a pre-trained language model to capture the deep contexts in news texts and an attention network to pool the output of PLM.
We denote an input news text with $M$ tokens as $[w_1, w_2, ..., w_M]$.
The PLM converts each token into its embedding, and then learns the hidden representations of words through several Transformer~\cite{vaswani2017attention} layers.
We denote the hidden token representation sequence as $[\mathbf{r}_1, \mathbf{r}_2, ..., \mathbf{r}_M]$.
We use an attention~\cite{yang2016hierarchical} network to summarize the hidden token representations into a unified news embedding.
The news embeddings learned by the PLM and attention network are further used for user modeling and candidate matching.

\subsection{Model Training}

Following~\cite{wu2019npa,wu2019nrms}, we also use negative sampling techniques to build labeled samples from raw news impression logs, and we use the cross-entropy loss function for model training by classifying which candidate news is clicked.
By optimizing the loss function via backward-propagation, the parameters in the recommendation model and PLMs can be tuned for the news recommendation task.

%However, many PLMs are in large size, and in news recommendation  the PLM needs to encode both candidate news and a sequence of clicked news.
%Thus, directly finetuning the entire PLM will lead to extremely heavy computational cost in model training.
%To handle this challenge, we design a partial model finetuning strategy to train the PLM-based news recommendation model effectively and efficiently.
%More specifically, motivated by the findings in~\cite{chi2020infoxlm}, we only use the first $M$ Transformer layers to generate the hidden word representations.
%This is because the top layers of PLMs may be adjusted to self-supervision tasks like masked word prediction, while the intermediate layers may encode more universal text information.
%In addition, since backward propagation is much more time-consuming than forward propagation, we only finetune the last $P$ layers of the partial model with $M$ layers to reduce the depth of backward propagation.
%In this way, the recommendation model can be trained in an efficient way.

%% file: data/experiment.tex
\section{Experiments}\label{sec:Experiments}

\subsection{Datasets and Experimental Settings}

Our offline experiments are conducted on two real-world datasets.
The first one is  MIND~\cite{wu2020mind}, which is an English dataset for monolingual news recommendation.
It contains the news click logs of 1 million users on Microsoft News in six weeks.\footnote{More details are on https://msnews.github.io/.}
The second one is a multilingual news recommendation dataset (denoted as \textit{Multilingual}) collected by ourselves on MSN News platform from Dec. 1, 2020 to Jan. 14, 2021.
It contains users from 7 countries with different language usage, and their market language codes are EN-US, DE-DE, FR-FR, IT-IT, JA-JP, ES-ES and KO-KR, respectively.
We randomly sample 200,000 impression logs in each market.
The logs in the last week are used for test and the rest are used for training and validation (9:1 split).
The detailed statistics of the two datasets are shown in Table~\ref{dataset}.

\begin{table}[h]
\centering
\caption{Detailed statistics of the two datasets.}\label{dataset}
%\resizebox{0.48\textwidth}{!}{
\begin{tabular}{lcc}
\Xhline{1.5pt}
                   & \textbf{MIND} & \textbf{Multilingual} \\ \hline
\# Users           & 1,000,000     & 1,392,531             \\
\# News            & 161,013       & 4,378,487             \\
\# Impressions     & 15,777,377    & 1,400,000             \\
\# Click Behaviors & 24,155,470    & 1,814,927             \\ \Xhline{1.5pt}
\end{tabular}
%}

\end{table}

In our experiments, we used the ``Base'' version of different pre-trained language models if not specially mentioned.
%Motivated by~\cite{chi2020infoxlm}, we used the first 8 layers of PLMs to generate text representations.
We finetuned the last two Transformer layers because we find there is only a very small performance difference between finetuning all layers and the last two layers.
Following~\cite{wu2020mind}, we used the titles of news for news modeling.
We used Adam~\cite{kingma2014adam} as the optimization algorithm and the learning rate was 1e-5.
The batch size was 128.\footnote{We used 4 GPUs in parallel and the batch size on each of them was 32.}
%The dropout~\cite{srivastava2014dropout} ratio was 20\%.
These hyperparameters are developed on the validation sets.
We used average AUC, MRR, nDCG@5 and nDCG@10 over all impressions as the performance metrics.
We repeated each experiment 5 times independently and reported the average performance.

\subsection{Offline Performance Evaluation}

We first compare the performance of several methods on the \textit{MIND} dataset to validate the effectiveness of PLM-based models in  monolingual news recommendation.
We compared several recent news recommendation methods including EBNR~\cite{okura2017embedding}, NAML~\cite{wu2019}, NPA~\cite{wu2019npa}, LSTUR~\cite{an2019neural}, NRMS~\cite{wu2019nrms} and their variants empowered by different pre-trained language models, including BERT~\cite{devlin2019bert}, RoBERTa~\cite{liu2019roberta} and UniLM~\cite{bao2020unilmv2}.
The results are shown in Table~\ref{table.performance}.
Referring to this table, we find that incorporating pre-trained language models can consistently improve the performance of basic models.\footnote{The results of t-test  show the improvements are significant ($p<0.001$).}
This is because pre-trained language models have stronger text modeling ability than  the shallow models learned from scratch in the news recommendation.
In addition, we find that the models based on RoBERTa are better than those based on BERT.
This may be because RoBERTa has better hyperparameter settings than BERT and is pre-trained on larger corpus for a longer time.
Besides, the models based on UniLM achieve the best performance.
This may be due to UniLM can exploit the self-supervision information in both text understanding and generation tasks, which can help learn a higher-quality PLM.

\begin{table}[!t]
 \caption{Performance of different methods on \textit{MIND}.} \label{table.performance} 
\begin{tabular}{lcccc}
\Xhline{1.5pt}
\textbf{Methods} & \textbf{AUC}   & \textbf{MRR}   & \multicolumn{1}{l}{\textbf{nDCG@5}} & \multicolumn{1}{l}{\textbf{nDCG@10}} \\ \hline
EBNR             & 66.54 & 32.43 & 35.38 & 40.09         \\
EBNR-BERT        & 69.56 & 34.77 & 38.04 & 43.72         \\
EBNR-RoBERTa     & 69.70 & 34.84 & 38.21 & 43.88         \\
EBNR-UniLM       & 70.56 & 35.31 & 38.65 & 44.32         \\ \hline
NAML             & 67.78          & 33.24          & 36.19             & 41.95          \\
NAML-BERT        & 69.42          & 34.66          & 37.91             & 43.65          \\
NAML-RoBERTa     & 69.60          & 34.78          & 38.13             & 43.79          \\
NAML-UniLM       & 70.50          & 35.26          & 38.60             & 44.27          \\ \hline
NPA              & 67.87 & 33.20 & 36.26 & 42.03        \\
NPA-BERT         & 69.50 & 34.72 & 37.96 & 43.72        \\
NPA-RoBERTa      & 69.64 & 34.81 & 38.14 & 43.82        \\
NPA-UniLM        & 70.52 & 35.29 & 38.63 & 44.29        \\ \hline
LSTUR            & 68.04 & 33.31 & 36.28 & 42.10        \\
LSTUR-BERT       & 69.49 & 34.72 & 37.97 & 43.70        \\
LSTUR-RoBERTa    & 69.62  & 34.80 & 38.15 & 43.79        \\
LSTUR-UniLM      & 70.56 & 35.29 & 38.67 & 44.31        \\ \hline
NRMS             & 68.18          & 33.29          & 36.31             & 42.20          \\
NRMS-BERT        & 69.50          & 34.75          & 37.99             & 43.72          \\
NRMS-RoBERTa     & 69.56          & 34.81          & 38.05             & 43.79          \\
NRMS-UniLM       & \textbf{70.64} & \textbf{35.39} & \textbf{38.71}    & \textbf{44.38} \\ \Xhline{1.5pt}
\end{tabular}
\end{table}

In addition, we conduct experiments on the \textit{Multilingual} dataset to validate the effectiveness of PLMs in multilingual news recommendation.
We compare the performance of EBNR, NAML, NPA, LSTUR and NRMS with different multilingual text modeling methods, including:
(1) MUSE~\cite{lee2017muse}, using modularizing unsupervised sense embeddings; 
(2) Unicoder~\cite{huang2019unicoder}, a universal language encoder pre-trained by cross-lingual self-supervision tasks; 
and (3) InfoXLM~\cite{chi2020infoxlm}, a contrastively pre-trained cross-lingual language model based on information-theoretic framework.
In these methods, following~\cite{huang2019unicoder} we mix up the training data in different languages.
In addition, we also compare the performance of independently learned monolingual models based on MUSE for each market (denoted as Single).
The results of different methods in terms of AUC are shown in Table~\ref{table.performance2}.
We find that multilingual models usually outperform the independently learned monolingual models.
This may be because different languages usually have some inherent relatedness and users in different countries may also have some similar interests.
Thus, jointly training models with multilingual data can help learn a more accurate recommendation model.
It also provides the potential to use a unified recommendation model to serve users in different countries with diverse language usage (e.g., Indo-European and Altaic), which can greatly reduce the computation and memory cost of online serving.
In addition, the performance methods based on multilingual PLMs are better than those based on MUSE embeddings.
This may be because PLMs are also stronger than word embeddings in capturing the complicated multilingual semantic information.
In addition, InfoXLM can better empower multilingual news recommendation than Unicoder.
This may be because InfoXLM uses better contrastive pre-training strategies than Unicoder to help learn more accurate models.

\begin{table}[!t]
 \caption{Performance of different methods on \textit{Multilingual}.} \label{table.performance2} 
 \resizebox{1.0\linewidth}{!}{
\begin{tabular}{lccccccc}
\Xhline{1.5pt} 
\textbf{Methods} & \textbf{EN-US} & \textbf{DE-DE} & \textbf{FR-FR} & \textbf{IT-IT} & \textbf{JA-JP} & \textbf{ES-ES} & \textbf{KO-KR} \\ \hline
EBNR-Single    & 62.08 & 59.94 & 61.66 & 60.27 & 61.57 & 58.30 & 63.53 \\
EBNR-MUSE      & 62.26 & 60.19 & 61.75 & 60.44 & 61.74 & 57.53 & 63.78 \\
EBNR-Unicoder  & 63.35 & 61.44 & 62.34 & 61.18 & 62.76 & 58.70 & 64.80 \\
EBNR-InfoXLM   & 64.29 & 62.03 & 62.97 & 61.98 & 63.34 & 59.33 & 65.58 \\ \hline
NAML-Single    & 62.05 & 59.89 & 61.56 & 60.21 & 61.54 & 58.21 & 63.5  \\
NAML-MUSE      & 62.17 & 60.17 & 61.71 & 60.4  & 61.69 & 57.46 & 63.73 \\
NAML-Unicoder  & 63.3  & 61.37 & 62.32 & 61.16 & 62.74 & 58.61 & 64.77 \\
NAML-InfoXLM   & 64.27 & 61.98 & 62.94 & 61.91 & 63.29 & 59.33 & 65.49 \\ \hline
NPA-Single     & 62.09 & 59.90 & 61.56 & 60.24 & 61.57 & 58.24 & 63.56 \\
NPA-MUSE       & 62.23 & 60.21 & 61.78 & 60.44 & 61.75 & 57.47 & 63.71 \\
NPA-Unicoder   & 63.32 & 61.41 & 62.35 & 61.20 & 62.77 & 58.64 & 64.80 \\
NPA-InfoXLM    & 64.29 & 62.00 & 62.93 & 61.94 & 63.31 & 59.37 & 65.50 \\ \hline
LSTUR-Single   & 62.09 & 59.95 & 61.58 & 60.22 & 61.58 & 58.22 & 63.57 \\
LSTUR-MUSE     & 62.21 & 60.21 & 61.79 & 60.44 & 61.73 & 57.49 & 63.75 \\
LSTUR-Unicoder & 63.34 & 61.40 & 62.36 & 61.20 & 62.77 & 58.65 & 64.80 \\
LSTUR-InfoXLM  & 64.31 & 62.03 & 62.96 & 61.95 & 63.32 & 59.38 & 65.54 \\ \hline
NRMS-Single    & 62.11 & 59.94 & 61.62 & 60.28 & 61.57 & 58.30 & 63.64 \\
NRMS-MUSE      & 62.33 & 60.29 & 61.86 & 60.54 & 61.90 & 57.62 & 63.93 \\
NRMS-Unicoder  & 63.41 & 61.50 & 62.46 & 61.22 & 62.81 & 58.84 & 64.79 \\
NRMS-InfoXLM   & \textbf{64.34}& \textbf{62.05}& \textbf{63.04}&\textbf{61.98}&\textbf{63.40}&\textbf{59.44}&\textbf{65.58}  \\ 
\Xhline{1.5pt} 
\end{tabular}
}
\end{table}

\begin{figure}[!t]
	\centering
	\includegraphics[width=0.28\textwidth]{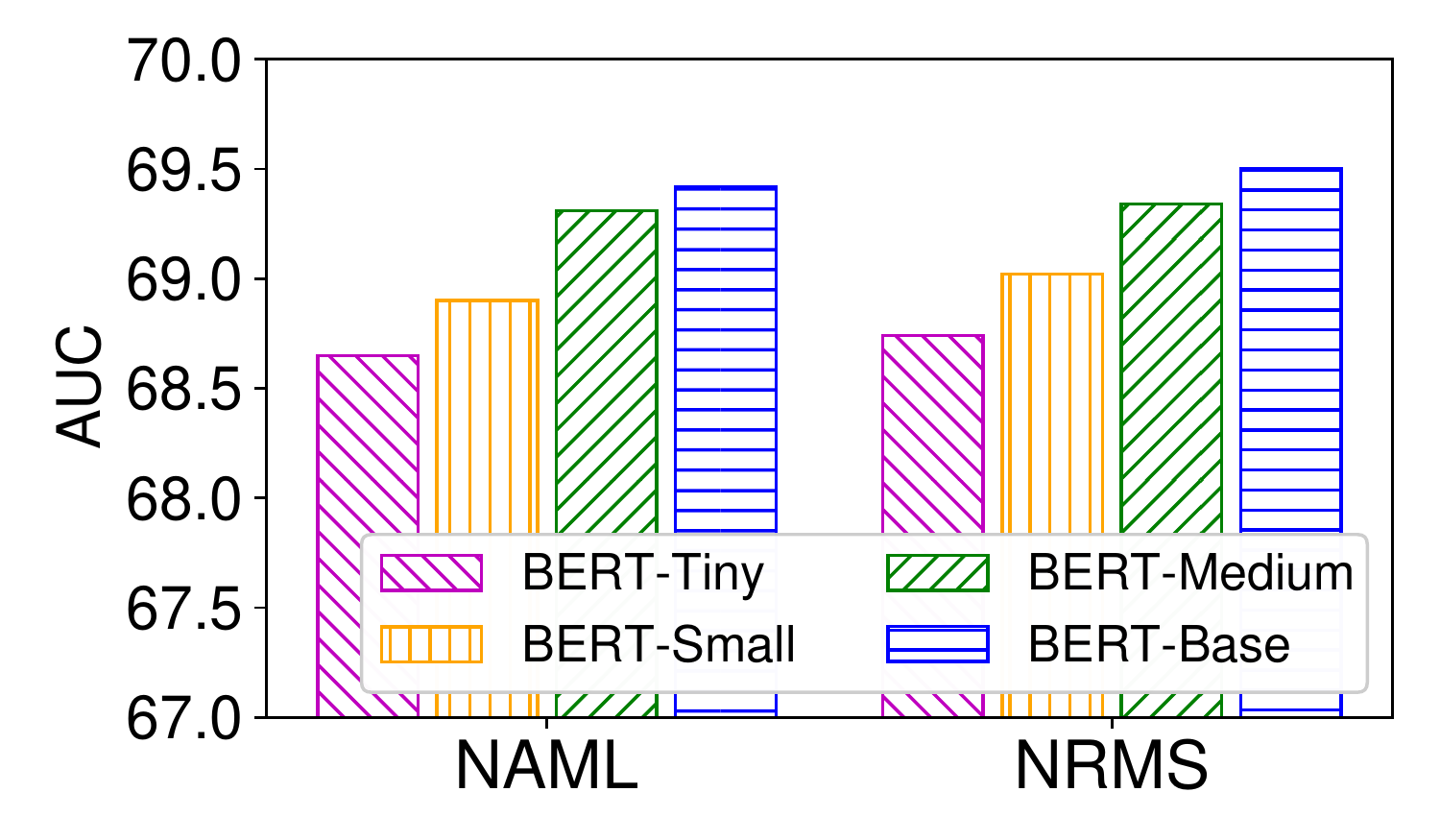}
	
\caption{Influence of the size of PLMs.}\label{fig.size}\vspace{-0.1in}

\end{figure}

\subsection{Influence of Model Size}

Next, we explore the influence of  PLM size on the recommendation performance.
We compare the performance of two representative methods  (i.e., NAML and NRMS) with different versions of BERT, including BERT-Base (12 layers), BERT-Medium (8 layers), BERT-Small (4 layers) and BERT-Tiny (2 layers).
The results on \textit{MIND} are shown in Fig.~\ref{fig.size}.
We find that using larger PLMs with more parameters usually yields better recommendation performance.
This may be because larger PLMs usually have stronger abilities in capturing the deep semantic information of news, and the performance may be further improved if more giant PLMs (e.g., BERT-Large) are incorporated.
However, since huge PLMs are too cumbersome for online applications, we prefer the base version of PLMs.

\begin{figure}[!t]
	\centering
	\includegraphics[width=0.28\textwidth]{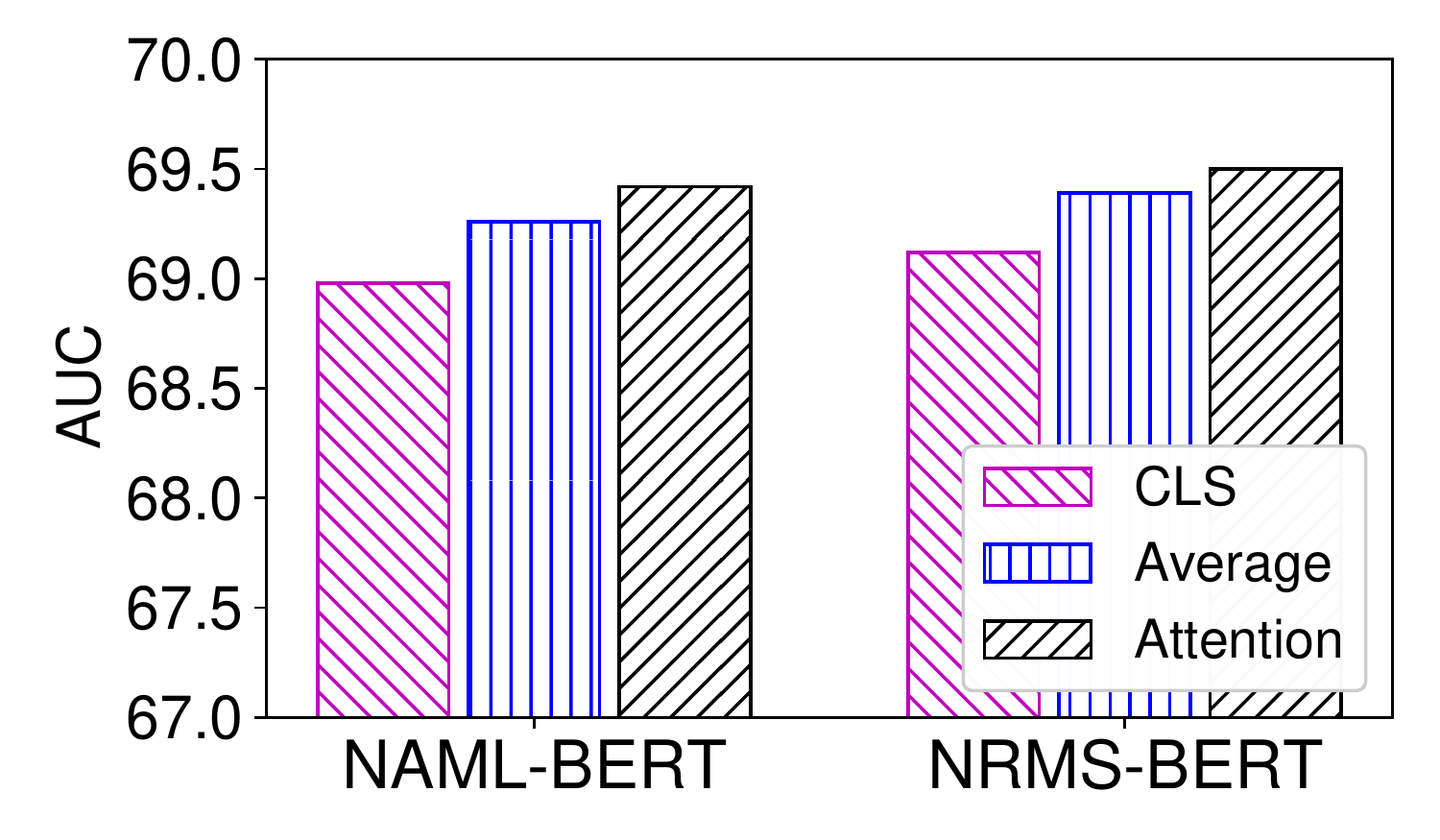}
\caption{Influence of the pooling methods.}\label{fig.pool}\vspace{-0.1in}
\end{figure}

\subsection{Influence of Different Pooling  Methods}

We also explore using different pooling methods for learning news embeddings from the hidden states of PLMs.
We compare three methods, including: (1) \textit{CLS}, using the representation of the ``[CLS]'' token as news embedding, which is a widely used method for obtaining sentence embedding; (2) \textit{Average}, using the average of hidden states of PLM; (3) \textit{Attention}, using an attention network to learn news embeddings from hidden states.
The results of NAML-BERT and NRMS-BERT on \textit{MIND} are shown in Fig.~\ref{fig.pool}.\footnote{We observe similar phenomenons in other PLM-based methods.}
We find it is very interesting that the \textit{CLS} method yields the worst performance.
This may be because it cannot exploit all output hidden states of the PLM.
In addition, \textit{Attention} outperforms \textit{Average}.
This may be because attention networks can distinguish the informativeness of hidden states, which can help learn more accurate news representations.
Thus, we choose attention mechanism as the pooling method.

\begin{figure}[!t]
	\centering
	\subfigure[NRMS.]{
	\includegraphics[width=0.22\textwidth]{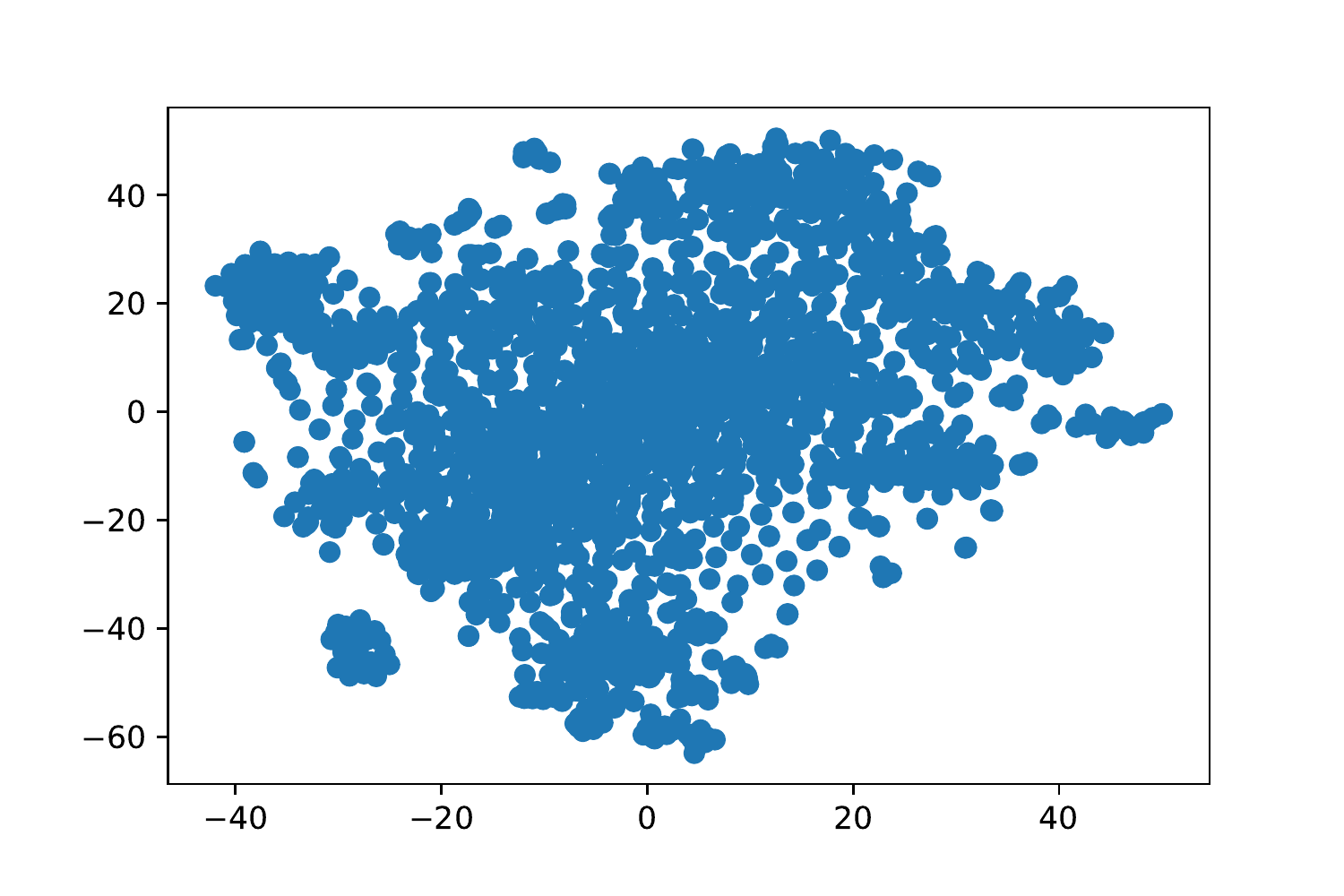}
	}
		\subfigure[NRMS-UniLM.]{
	\includegraphics[width=0.22\textwidth]{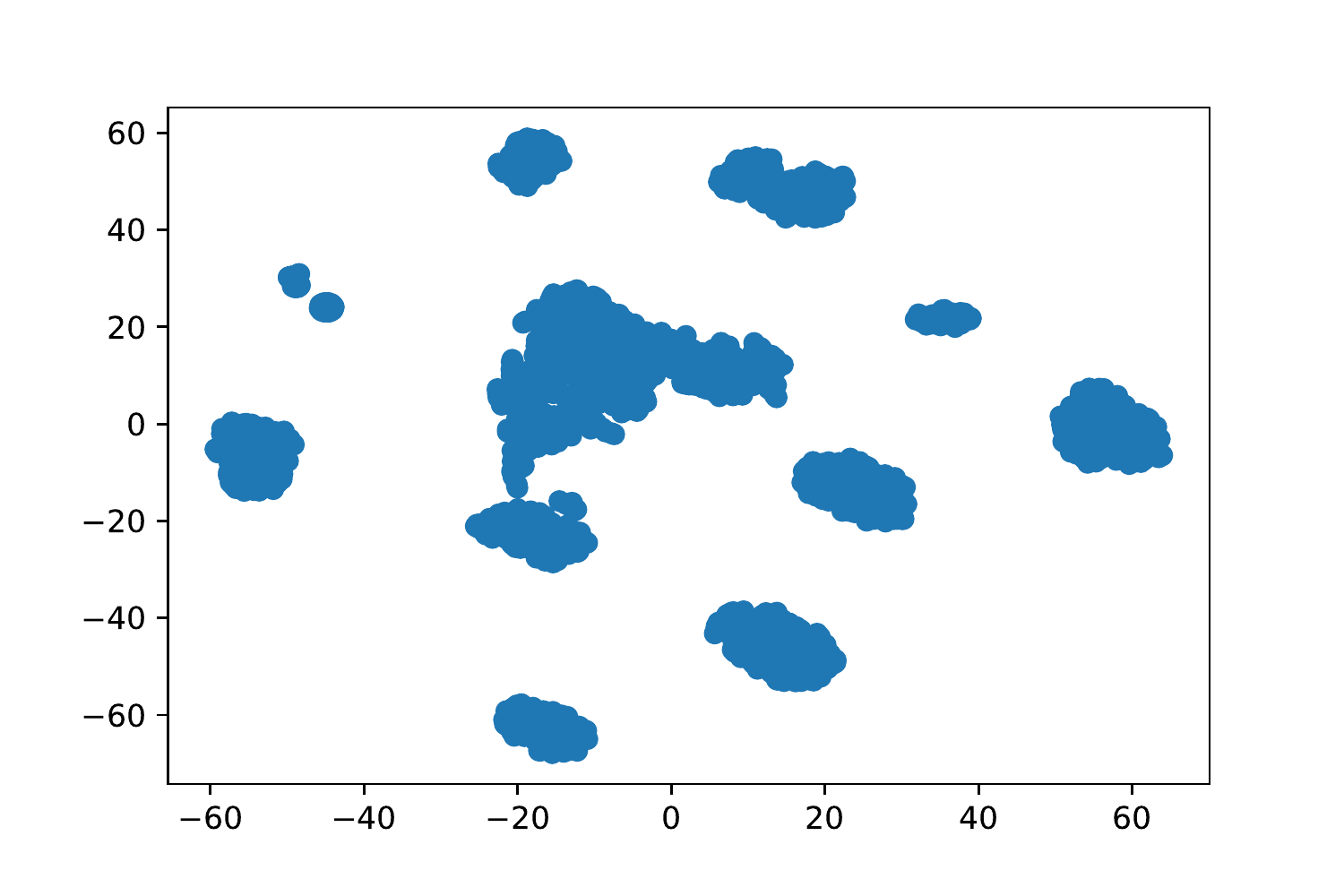}
	}
\caption{Visualization of news embeddings learned by NRMS and NRMS-UniLM.}\label{fig.vis}
\vspace{-0.1in}
\end{figure}

\subsection{Visualization of News Embedding}

We also study the differences between the news embeddings learned by shallow models and PLM-empowered models. 
We use t-SNE~\cite{van2008visualizing} to visualize the news embeddings learned by NRMS and NRMS-UniLM, and the results are shown in Fig.~\ref{fig.vis}.
We find an interesting phenomenon that the news embeddings learned by NRMS-UniLM are much more discriminative than NRMS. 
This may be because the shallow self-attention network in NRMS cannot effectively model the semantic information in news texts.
Since user interests are also inferred from embeddings of clicked news, it is difficult for NRMS to accurately model user interests from non-discriminative news representations.
In addition, we observe that the news embeddings learned by NRMS-UniLM form several clear clusters.
This may be because the PLM-empowered model can disentangle different kinds of news for better user interest modeling and news matching.
These results demonstrate that deep PLMs have greater ability than shallow NLP models in learning discriminative text representations, which is usually beneficial for accurate news recommendation.

\subsection{Online Flight Experiments}

We have deployed our PLM-empowered news recommendation models into the Microsoft News platform.
Our NAML-UniLM model was used to serve users in English-speaking markets, including EN-US, EN-GB, EN-AU, EN-CA and EN-IN.
The online flight experimental results have shown a gain of 8.53\% in click and 2.63\% in pageview against the previous news recommendation model without pre-trained language model.
In addition, our NAML-InfoXLM model was used to serve users in other 43 markets with different languages.
The online flight results show an improvement of 10.68\% in click and 6.04\% in pageview.
These results validate that incorporating pre-trained language models into news recommendation can effectively improve the recommendation performance and user experience of online news services.

%% file: data/conclusion.tex
\section{Conclusion}\label{sec:Conclusion}

In this paper, we present our work on empowering personalized news recommendation with  pre-trained language models.
We conduct extensive offline experiments on both English and multilingual news recommendation datasets, and the results show incorporating pre-trained language models can effectively improve news modeling for news recommendation.
In addition, our PLM-empowered news recommendation models have been deployed to a commercial news platform, which is the first public reported effort to empower real-world large-scale news recommender systems with PLMs.
The online flight results show significant improvement in both click and pageview in a large number of markets with different languages.